\pdfoutput=1
\documentclass[conference,compsoc]{IEEEtran}
\usepackage[T1]{fontenc}
\usepackage{microtype}
\usepackage{amsmath,amssymb}
\usepackage{booktabs,makecell,multirow}
\usepackage{graphicx}
\usepackage{xcolor}
\usepackage[numbers,sort&compress]{natbib}
\usepackage{xurl}
\usepackage{adjustbox}
\usepackage{placeins}
\usepackage{fancyvrb}
\usepackage{tikz}
\usepackage[hidelinks]{hyperref}
\usepackage[capitalize,noabbrev]{cleveref}

\newcommand{\hesp}{HESP}

\DeclareRobustCommand{\stepnum}[1]{\tikz[baseline=-0.6ex]\node[circle,fill=black,text=white,font=\scriptsize\bfseries,inner sep=0.6pt,minimum size=9pt]{#1};}
\graphicspath{{figures/}}
\renewcommand{\paragraph}[1]{\par\vspace{0.6ex}\noindent\textbf{#1.}\hspace{0.45em}\ignorespaces}

\begin{document}

\title{\hesp{}: Separating What to Probe from When to Stop\\in Local LLM Alert-Triage Agents}

\author{\IEEEauthorblockN{Zhuowen Liu}
\IEEEauthorblockA{Cybersecurity Lab\\
Japan Advanced Institute of\\ Science and Technology (JAIST)\\
Nomi, Ishikawa, Japan\\
\texttt{s2410431@jaist.ac.jp}}
\and
\IEEEauthorblockN{Zhixuan Wang\textsuperscript{*}}
\IEEEauthorblockA{College of Fine Arts\\
Liaoning Normal University\\
Dalian, Liaoning, China\\
\texttt{wangzhixuanwzx@gmail.com}\\
\textsuperscript{*}Corresponding author}}

\maketitle

\begin{abstract}
Security operations centers receive far more alerts than analysts can investigate, and organizations that cannot send their telemetry to hosted models must automate triage with small open-weight LLMs on their own hardware. Current LLM agents leave the investigation procedure to the model, and small local models fail at it: they probe without converging, never commit to a verdict, or dismiss real attacks. In this paper, we present \hesp{}, a controller that holds the investigation procedure outside the model. \hesp{} keeps a ledger of competing explanations, selects read-only probes by expected information gain per cost, accepts only verdicts backed by current evidence, can end an investigation itself, and journals every prediction before its observation. We evaluated \hesp{} in four pre-registered studies with five open-weight models from two families (7B to 72B), totalling 7{,}272 audited episodes in a controlled triage environment. With likelihood tables counted from LLM-free runs, \hesp{} lifts Qwen2.5-7B from 0.125 to 1.000 verified completion, matching oracle tables. The information-gain ranking adds $+0.26$ to $+0.35$ on every model that concludes, and a controller-side stop lifts Llama-3.1-8B, which never concludes on its own, from 0 to 0.917. What to probe and when to stop are therefore separate failures, and different small models exhibit different ones. Because \hesp{} and its planner run entirely on local hardware, it suits environments where telemetry cannot leave the premises. We release all code, protocols, and episode journals.
\end{abstract}

\section{Introduction}
\label{sec:intro}

When an alert fires, an analyst has to decide what caused it: an attack in progress, a misconfiguration, authorized activity, or a false positive. The analyst narrows the candidate explanations by querying authentication logs, request patterns, configuration, change tickets, and threat intelligence, and stops once some observation supports one explanation and the others have been ruled out. The verdict decides whether the incident is escalated, and every query costs time.

Doing this by hand does not scale. Alert volumes exceed what analysts can investigate, and most alerts turn out to be benign~\citep{nodoze}. LLM agents that choose queries and read their results fit this loop naturally. Many organizations, however, cannot or will not send raw telemetry to a hosted model. For them the realistic option is an open-weight model small enough to run on a single local GPU, and recent work on LLM-driven attack investigation already reports that such models fall behind proprietary ones on complex cases~\citep{clouseau}.

Efforts to automate triage and investigation follow two broad approaches. The first builds provenance graphs of host activity and ranks or summarizes them with fixed rules or learned models~\citep{nodoze,kairos}. These systems reduce what an analyst has to read, but they do not decide which query to run next. The second applies LLM agents that choose queries and interpret the results~\citep{excytin,clouseau}, in the ReAct pattern where the model decides every step~\citep{react}. These agents leave the investigation procedure to the model, and with small local models the procedure is exactly what fails. In our pre-registered studies, Qwen2.5-7B choosing its own probes ends only 8 to 13\% of triage cases with a verified verdict: it keeps probing until its budget runs out. Llama-3.1-8B never concludes. Across 3{,}389 decisions it asks for one more probe every time, even when the evidence is overwhelming and no legal probe is left. Larger models do conclude, but without guidance they cite invalid evidence in 22 to 31\% of their citations and call 25 to 37\% of actionable incidents benign.

In this paper, we introduce \hesp{}, a controller that holds the investigation procedure outside a local LLM agent. An investigation consists of two decisions, \emph{what to probe next} and \emph{when the evidence suffices}, and a controller can own either or both. \hesp{} keeps a ledger of competing explanations, updated by Bayes' rule from tables that state what each probe returns under each explanation. It ranks read-only probes by expected information gain per unit cost, accepts only verdicts backed by a current supporting observation, and can end the investigation itself. The model remains the planner: it reads the ledger and may propose probes, finish, or abstain. Every prediction is journaled before its observation, and every episode is audited afterwards.

We evaluated \hesp{} in four pre-registered studies with five open-weight models from two families (7B to 72B) and 7{,}272 audited episodes in a controlled triage environment. Tables counted from LLM-free development episodes lift Qwen2.5-7B from 0.125 to 1.000 verified completion, matching tables taken from the environment's generator. With the planner's information held fixed, the information-gain ranking adds $+0.26$ to $+0.35$ on every model that concludes. A controller-side stop lifts Llama-3.1-8B from 0 to 0.917, after which controller probe selection adds no detectable completion for it, whereas Qwen2.5-7B needs both parts. The controller alone, with no LLM, completes 0.917 of cases, and the best LLM configurations reach 1.000 because strong models wait for confirming evidence before they finish.

In summary, the contributions of this paper are:
\begin{itemize}
  \item We show that small local LLMs fail at alert triage on procedure, and that choosing what to probe and deciding when to stop are separate failures that differ between models of similar size (\S\ref{sec:motivation}, \S\ref{sec:rq-stop}).
  \item We design, implement, and open-source \hesp{}\footnote{Code, frozen protocols, prediction tables, outcome files, and episode journals are available at \url{https://github.com/lzwhehe/HESP}.}, a controller that owns the hypothesis ledger, probe selection, verdict acceptance, and stopping for a local LLM planner, with a write-ahead journal and a per-episode audit (\S\ref{sec:approach}).
  \item We conduct four pre-registered studies with five open-weight models, using frozen protocols and source hashes, blind controls that hold the planner's prompt fixed across selectors, and a factorial design that separates probe selection from stopping (\S\ref{sec:setup}, \S\ref{sec:results}).
  \item We show that the controller needs no oracle: likelihood tables counted from a few LLM-free runs work as well as tables derived from the environment's generator, whereas tables elicited from the model do not (\S\ref{sec:rq-help}).
  \item We measure three public security datasets and show why none of them can test an investigation policy, and we present a case study on recorded attack telemetry of the kind of case they lack (\S\ref{sec:realdata}).
\end{itemize}

\section{Background}
\label{sec:background}

\subsection{Alert Triage}
Alert triage follows detection: for each alert, the analyst determines what actually happened and whether a response is needed~\citep{nodoze}. Attack investigation reconstructs a complete attack story from a point of interest~\citep{kairos,clouseau}; triage instead ends with a single verdict, namely the cause of the alert together with the evidence that supports it. The analyst reaches the verdict by querying the environment. Queries differ in cost and in how well they separate the candidate explanations. Some are cheap and decisive for one explanation. Others, such as a threat-intelligence lookup, are slow and return the same answer for several attacks. The order of queries therefore determines how quickly an investigation ends, and the analyst must also judge when the evidence already suffices.

Consider an alert that reports a spike in authentication failures and HTTP errors on a web tier. It looks like credential stuffing, but the same symptom can come from SQL-injection probing, an authorized vulnerability scan, a health check misclassified as an attack, or an administration panel exposed to the public network. A threat-intelligence lookup on the top source may call it malicious, an answer that three of these causes share. Reading the admin panel's configuration, one cheap query, settles the last explanation at once, but only an investigator who keeps that explanation in view will run it.

\subsection{Local LLM Agents}
An LLM agent couples a language model with tools. At each step the model reads the task and the observations so far, then either calls a tool or ends the task~\citep{react}. Extensions add verbal self-reflection~\citep{reflexion} or learn when to call tools~\citep{toolformer}. In security, such agents have been applied to penetration testing~\citep{pentestgpt}, capture-the-flag challenges~\citep{cybench}, and investigation over security logs~\citep{excytin,clouseau}, mostly with large hosted models. Open-weight models such as the Qwen2.5 and Llama~3.1 families can instead run on an organization's own hardware; with 4-bit weight quantization, models of up to 72B parameters fit on a single GPU. Local deployment keeps telemetry on the premises, but it trades away model strength, and in an agent loop every step's decision rests on that weaker model.

\section{Motivation}
\label{sec:motivation}

\subsection{Challenges}
When a small local model drives the investigation, three challenges arise. We measured each on the baseline configurations of our studies (\S\ref{sec:setup}), in which the model chooses its own probes.

\stepnum{1} \textbf{Probing without converging.} Qwen2.5-7B selecting its own probes verifies only 0.083 to 0.125 of cases. It runs about seven distinct probes per episode, spends 9.4 to 9.5 of its 10 cost units, and issues a verdict in only 6 or 7 of 72 episodes; three quarters of its episodes end with the tool budget exhausted. Replacing its choices with uniformly random ones raises completion by $+0.556$ (\S\ref{sec:rq-rank}), so its own choices are worse than chance.

\stepnum{2} \textbf{Never deciding to stop.} Llama-3.1-8B requests another probe in every one of its 3{,}389 decisions, including those in which the ledger's posterior exceeds 0.999 and the prompt states that no legal probe is left. Its outputs are valid JSON throughout; it simply never finishes. Better probing cannot help a model that never states a verdict.

\stepnum{3} \textbf{Verdicts without support.} Larger models conclude, but not always on evidence. Without a ledger, Qwen2.5-32B and 72B call 25\% and 37\% of actionable incidents benign and cite invalid evidence in 22\% and 31\% of their citations (\S\ref{sec:security}). In triage, a verdict that dismisses a real attack is the costliest error, because the incident is closed without a response.

\subsection{Our Solution}
We now describe how \hesp{} addresses these challenges.

\stepnum{1} \textbf{Information-gain selection outside the model.} \hesp{} ranks every legal probe by the expected reduction in uncertainty over the candidate causes per unit cost, and executes the best one whatever the model proposed. The likelihoods behind the ranking are counted from LLM-free development runs. They are not elicited from the model, which cannot supply them: elicited tables leave completion at 0.069 to 0.375 (\S\ref{sec:rq-help}).

\stepnum{2} \textbf{A controller-side stop.} Before each planner call, \hesp{} applies an explicit acceptance rule. Once the leading cause has posterior at least 0.8 and a current observation supports it, the controller concludes on its own and cites that evidence. The model may still finish earlier, so a strong model keeps the option of waiting for confirmation.

\stepnum{3} \textbf{Evidence-bound verdicts.} \hesp{} accepts a verdict only if it cites a current observation that raised the claimed cause's score. An independent verifier, which the planner cannot reach, checks each verdict against a signature that the cause alone produces, and a write-ahead journal lets an audit replay every step.

\section{Approach}
\label{sec:approach}

\begin{figure*}[t]
  \centering
  \includegraphics[width=\textwidth]{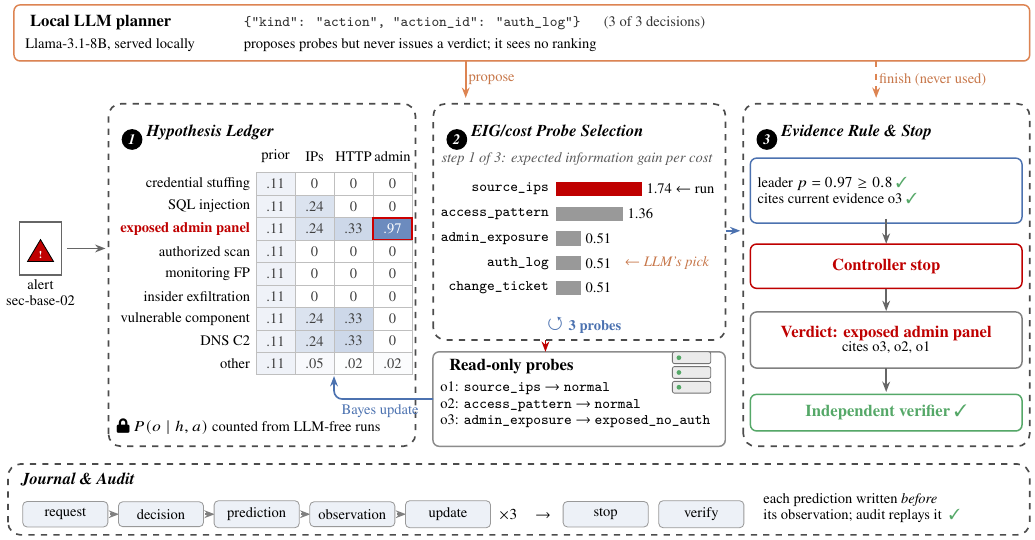}
  \caption{Overview of \hesp{}, drawn on the running example of this section (Llama-3.1-8B as planner; the true cause is an exposed admin panel). \protect\stepnum{1} The hypothesis ledger keeps a posterior over the candidate causes, updated by Bayes' rule from frozen probe-outcome tables. \protect\stepnum{2} Before each step the controller ranks the legal probes by expected information gain per cost and runs the best one; the planner's proposal (here \texttt{auth\_log}, three times) is advisory. \protect\stepnum{3} Once the leader's posterior reaches 0.8 and a current observation supports it, the controller concludes and cites that evidence, and an independent verifier checks the verdict. Every step is journaled before its observation and audited afterwards. All values are taken from the archived journal.}
  \label{fig:pipeline}
\end{figure*}

\hesp{} (Hypotheses $\rightarrow$ Evidence $\rightarrow$ State $\rightarrow$ Planning) is a controller that sits between a local LLM planner and a catalogue of read-only probes (\cref{fig:pipeline}). The planner proposes; the controller decides which probe runs, which verdict is accepted, and when the investigation ends. The workflow has two phases:
\begin{enumerate}
  \item \emph{Table construction:} before deployment, LLM-free development runs execute every probe on cases with known causes, and \hesp{} counts the outcomes into prediction tables (\S\ref{sec:tables}).
  \item \emph{Investigation:} for each alert, \hesp{} maintains a hypothesis ledger (\S\ref{sec:ledger}), selects the probes to run (\S\ref{sec:selection}), and applies an evidence rule and a stopping rule to every verdict (\S\ref{sec:finish}). The planner reads the ledger and proposes decisions. Every step is written to a journal before its observation and audited afterwards.
\end{enumerate}

\paragraph{Scope and Assumptions} \hesp{} runs inside the organization's network, with a local open-weight model as the planner and read-only access to security telemetry through a fixed probe catalogue: authentication and HTTP logs, source reputation, DNS logs, configuration, inventory, change tickets, and threat intelligence. Verdicts go to a human analyst; the agent never remediates and never sends traffic to the systems under investigation. The alert may be caused by an attacker who controls part of the evidence the agent reads, such as request payloads, user agents, or command lines, and who wants the investigation to end with a benign verdict or with none. We trust the controller, its tables, the probe implementations, and the journal, and we do not trust the planner to follow a sound procedure. Three properties then hold whatever the planner outputs: (i) every executed probe is in the read-only catalogue, is affordable, and has not already run in the current state; (ii) every accepted verdict cites at least one current observation that raised the claimed cause's score; and (iii) every prediction is journaled before the observation it predicts. \hesp{} does not guarantee a correct verdict, which depends on the tables and on the hypothesis set containing the true cause. Evidence manipulation, prompt injection through log content, and the generation of new hypotheses fall outside the scope of this paper.

\subsection{Prediction Tables}
\label{sec:tables}
The ledger and the selector both need, for every probe $a$ and cause $h$, a distribution $P(o \mid h, a)$ over the probe's outcome vocabulary. \hesp{} estimates these tables by counting. An exhaustive script, with no LLM involved, runs every probe $k$ times on each development instance of every cause, records the outcomes, and normalizes the counts with $\varepsilon = 0.01$ smoothing. The tables are written to disk, hashed, and frozen before any evaluation episode runs, and a unit test enforces that the estimator never calls the environment's generating function. A deployment would take its development cases from resolved tickets or replayed incidents (\S\ref{sec:discussion}).

Asking the planner for the likelihoods would remove the need for development data, but small models cannot supply them (\S\ref{sec:rq-help}). Counting, in contrast, works with as few as five episodes per case in our environment.

\paragraph{Example} In the counted table used by our running example ($k{=}20$), the probe \texttt{admin\_exposure} returns \textit{exposed, no auth} with probability 0.995 when the admin panel is exposed and 0.005 under each of the other seven named causes. The probe \texttt{source\_ips} returns \textit{many residential} with probability 0.992 under credential stuffing and 0.002 under each of the other named causes.

\subsection{Hypothesis Ledger}
\label{sec:ledger}
An episode starts from one alert and a finite set $\mathcal{H}$ of candidate causes, which includes an explicit \textit{other}. The agent may run read-only probes $a \in \mathcal{A}$, each with cost $c(a)$ and an outcome $o$ from a finite vocabulary. Budgets bound the number of probes, their total cost, the number of planner decisions, and wall time. The episode ends with a verdict $(h, E)$, where $E$ is a set of cited observations, with an abstention, or with an exhausted budget. A verdict is correct if $h$ is the cause in effect and $E$ contains a current observation that this cause alone produces.

The ledger holds a posterior $p(h)$ over $\mathcal{H}$, starting from a uniform prior. After observing outcome $o$ of probe $a$, it applies Bayes' rule with the frozen table and records, for every cause, whether the observation raised, lowered, or left unchanged its score (\textit{support}, \textit{against}, \textit{neutral}). A transient failure, such as an HTTP 503, is logged but is not evidence, and the probe may be retried. An outcome that is impossible under every cause marks a table conflict and is skipped. The environment exposes a \emph{state version}. When the incident changes under the agent, the version advances and the current scores are reset to the prior; earlier evidence stays in the record but no longer counts as current.

\paragraph{Example} The alert of our running example reports a spike in authentication failures and HTTP errors, which leaves nine candidates at 0.111 each. The first probe, \texttt{source\_ips}, returns \textit{normal}, and four causes remain plausible at 0.238 each. The second, \texttt{access\_pattern}, also returns \textit{normal} and leaves three at 0.327: the exposed admin panel, a vulnerable component, and DNS command-and-control. The third, \texttt{admin\_exposure}, returns \textit{exposed, no auth} and moves the exposed admin panel to 0.966.

\subsection{Probe Selection}
\label{sec:selection}
For every legal probe, meaning one that is untried in the current state, affordable, and has its prerequisites met, the controller computes the expected information gain
\begin{align*}
\mathrm{EIG}(a) &= H(p) - \textstyle\sum_{o} P(o \mid a)\, H\big(p(\cdot \mid o, a)\big),\\
P(o \mid a) &= \textstyle\sum_h p(h)\, P(o \mid h, a),
\end{align*}
where $H$ is Shannon entropy and $p(\cdot \mid o, a)$ is the posterior after observing $o$. It ranks the probes by $\mathrm{EIG}(a)/c(a)$ and executes the top one, whatever the planner proposed.

\paragraph{Planner Design} At each step the planner receives a structured request rendered as text: the task, the candidate causes, the probe catalogue with costs and outcome descriptions, the observations so far, the ledger, feedback on blocked proposals, and the remaining budgets (Appendix~\ref{app:prompt} reproduces one verbatim). It returns one JSON decision: \texttt{action} (propose a probe), \texttt{finish} (a cause plus the IDs of the observations it cites), or \texttt{stop} (abstain). An invalid reply is returned to the model with the error for one retry. The configurations we evaluate differ in whether the planner is shown the controller's ranking. The \emph{blind} configurations state only that the controller picks the probe and show neither the ranking nor the selector, so the prompt is identical whichever selector runs, which a unit test checks. A \emph{random} selector, which picks uniformly among legal probes, serves as the control.

\paragraph{Example} In the running example, Llama-3.1-8B proposed \texttt{auth\_log} at each of its three decisions. The ranking valued that probe at 0.51 bits per cost unit before the first step, against 1.74 for \texttt{source\_ips}, and at 0.04 before the third, when \texttt{admin\_exposure} led with 0.86. The controller ran \texttt{source\_ips}, \texttt{access\_pattern}, and \texttt{admin\_exposure}.

\subsection{Evidence Rule, Stop, and Verification}
\label{sec:finish}
\paragraph{State-guarded finish} A planner's \texttt{finish} is accepted only if the claimed cause has current score at least $\tau = 0.8$ and at least one cited observation is current-state evidence that supports it. A rejected finish is returned to the planner as feedback.

\paragraph{Controller-side stop} The controller may also conclude on its own. Before each planner call, and when a budget runs out, it applies the same acceptance rule to the current leader. If the rule is met, it cites up to three supporting current-state observations and ends the episode. The planner may still finish or abstain earlier, and nothing in its prompt reveals the rule.

\paragraph{Independent verifier} A verdict counts as \emph{verified} only if the cause equals the cause in effect and at least one cited current-state observation carries a signature that this cause alone produces. A verdict reached purely by elimination, or backed only by a generic ``malicious'' threat-intelligence hit that several attacks share, does not verify. The planner has no access to the verifier.

\paragraph{Journal and audit} Every planner request and decision, every prediction, observation, and ledger update, and the verification are appended to a journal, and each prediction is written before the probe it predicts runs. After each episode, an audit replays the journal and checks event order, observation provenance, counters, budgets, citations, and that every verdict the controller made met its rule.

\paragraph{Example} Before the fourth planner call of the running example, the exposed admin panel led at 0.966 and the third observation supported it, so the controller concluded and cited all three observations. The verifier confirmed the verdict, because the third observation carries the exposed panel's own signature. The episode used three probes and three cost units and took 5.1\,s. With the same controller selection but without the stop, the same model reached the same observation at step three and never concluded (\cref{fig:motivating}, second row).

\section{Evaluation Setup}
\label{sec:setup}

To assess \hesp{}, we structure our evaluation around four questions:
\begin{enumerate}
  \item \emph{Reliability:} Does \hesp{} make small local models reliable triage agents, and does it depend on oracle knowledge of the environment? (\S\ref{sec:rq-help})
  \item \emph{Ranking versus Choice:} Is the gain due to the information-gain ranking, or merely to taking the choice of probe away from the model? (\S\ref{sec:rq-rank})
  \item \emph{Probing versus Stopping:} Are choosing what to probe and deciding when to stop separate failures, and which one does each model have? (\S\ref{sec:rq-stop})
  \item \emph{Failure Modes:} How do verdicts fail, in security terms? (\S\ref{sec:security})
\end{enumerate}
Each question was the subject of a pre-registered study designed after the previous one had run (\cref{tab:studies}). \S\ref{sec:realdata} then asks whether public security data could test the same questions.

\begin{table*}[t]
  \caption{Overview of the four pre-registered studies on sec-triage, 7{,}272 episodes in total. The repository labels the studies by version.}
  \label{tab:studies}
  \centering\small
  \begin{adjustbox}{max width=\textwidth}
  \begin{tabular}{llllrl}
    \toprule
    Study & Version & Question & Models & Episodes & Primary endpoint(s) \\
    \midrule
    Initial & v0.5 & does the controller help? & 3 Qwen & 1{,}944 & full \hesp{} vs.\ Memory-only, per model \\
    Table-source & v0.6 & must the tables come from an oracle? & 3 Qwen & 1{,}728 & counted-table \hesp{} vs.\ Memory-only, 7B \\
    Decomposition & v0.8 & ranking or loss of choice? second family & 5 & 1{,}800 & ranking vs.\ random (Q-7B); \hesp{} vs.\ Memory-only (L-8B) \\
    Stopping & v0.9 & what to probe vs.\ when to stop & 5 & 1{,}800 & controller stop; selection given stop (both L-8B) \\
    \bottomrule
  \end{tabular}
  \end{adjustbox}
\end{table*}

\subsection{Environment and Tables}

\subsubsection{The sec-triage Environment} Each episode is one alert on a web service. The hidden cause is one of eight explanations: three attacks (credential stuffing, SQL-injection probing, DNS command-and-control), an insider exfiltrating data, an exposed admin panel, a known-vulnerable component, an authorized vulnerability scan, and a monitoring false positive. The hypothesis set adds \textit{other}. Ten read-only probes, costing 1 to 3 units each, cover authentication failures, HTTP patterns, source reputation, admin exposure, per-user egress, DNS characteristics, component inventory, change tickets, threat intelligence, and a generic runbook. We designed the environment so that shortcuts fail. Threat intelligence says only that an indicator is malicious, and three attacks share that answer. The runbook carries no information. Several benign causes look like attacks until the right probe runs. Each cause appears in three variants. In \textit{base}, outcomes are deterministic except for a 10\% threat-intelligence lag. \textit{Noise} adds 15\% transient HTTP 503 failures and a 30\% lag. In \textit{drift}, the incident is superseded after the second probe: the state version advances and the true cause switches. This gives 24 tasks, and each study runs every task three times per configuration. Each episode is served from its own loopback HTTP sandbox.

\subsubsection{Prediction Tables} We compare three sources of $P(o \mid h, a)$. \emph{Designer} tables come from the environment's generating function and serve as an oracle reference. \emph{Counted} tables (``empirical'' in the tables) are estimated as in \S\ref{sec:tables}, with $k \in \{1, 5, 20, 100\}$ development episodes per cause and variant, from the base and noise variants only and with seeds drawn from a space disjoint from evaluation. \emph{Elicited} tables are produced by the planner LLM itself. Every table is frozen and hashed before any evaluation episode runs. Because development and evaluation episodes come from the same generator, the counted tables tell us how many observations are enough, not how the method behaves in an unfamiliar environment.

\begin{figure*}[t]
  \centering
  \includegraphics[width=\textwidth]{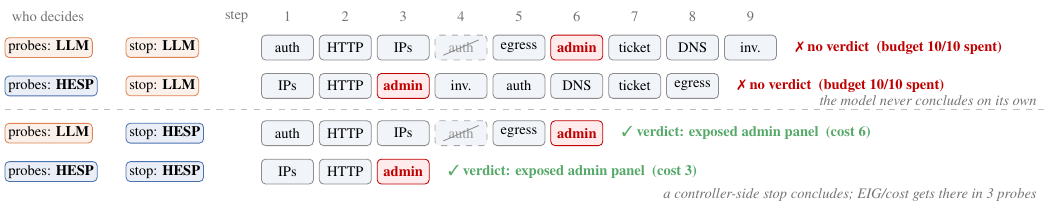}
  \caption{One alert, one local model, four ways to split the investigation. The alert's true cause is an exposed admin panel; the planner is Llama-3.1-8B in every row, with the same seed. Each chip is a probe that ran (dashed: a duplicate proposal the controller blocked); red marks the observation that confirms the cause. When the model decides when to stop (first two rows), it reaches the decisive evidence and keeps probing until the budget is gone. A controller-side stop concludes (last two rows), and information-gain selection gets there in three probes. All four rows are taken verbatim from the stopping study's archived journals.}
  \label{fig:motivating}
\end{figure*}

\subsection{Experiment Design}

\subsubsection{Configurations} All configurations use the same model, prompt template, probe catalogue, budgets (10 tool calls, 10 cost units, 12 decisions, 900\,s), duplicate blocking, and verifier. \emph{ReAct-style}: the planner sees the raw observation history only and picks every probe. \emph{Memory-only}: the planner also sees the ledger and still picks every probe. \emph{\hesp{}}: the controller picks the probe (EIG/cost or random), with or without showing its ranking, and with or without the finish guard and the controller-side stop. \Cref{fig:motivating} shows four of these configurations on the same alert. After the stopping study we also ran the frozen controller with the LLM replaced by a script that proposes probes in catalogue order and never finishes. This \emph{LLM-free reference} uses the stopping study's seeds, tasks, budgets, and table (216 episodes); it was not pre-registered and we report it as exploratory.

\subsubsection{Implementation} \hesp{} is about 2{,}900 lines of Python that use only the standard library. Planners are pluggable: one JSON decision protocol serves vLLM and Ollama endpoints and the scripted policies used for LLM-free runs. Each episode gets its own loopback HTTP sandbox, so probes are real HTTP requests with transient failures and state changes, and no episode can observe another. Every study uses Qwen2.5-7B-Instruct (bf16), Qwen2.5-32B-Instruct-AWQ, and Qwen2.5-72B-Instruct-AWQ; the decomposition and stopping studies add Llama-3.1-8B-Instruct (bf16) and Llama-3.1-70B-Instruct-AWQ-INT4. Models are served by vLLM on a single RTX~6000-class GPU at temperature 0.2, with a distinct seed per call. A paired, shuffled, resumable runner executes up to 32 episodes concurrently and refuses to resume if the manifest or the hash of the controller source has changed. After each model, it audits all episodes, archives the journals, and records the archive's SHA-256. The stopping study's 1{,}800 episodes ran in 28 minutes. A suite of 145 unit tests covers the ledger, selectors, blind prompts, guard, controller stop, audit, and the isolation of the table estimator from the environment's generator.

\subsubsection{Evaluation Methodology} Each study's hypotheses, configurations, primary endpoints, and analysis were written down and frozen, together with the source and table hashes, before any of its episodes ran; changes after freezing were appended as errata (Appendix~\ref{app:prereg}). The outcome measure is \emph{verified completion}, the fraction of episodes that end with a verified verdict. Every other ending, including timeouts and malformed outputs, counts as a failure. We report paired per-task differences with percentile task-cluster bootstrap intervals (2{,}000 resamples, 24 clusters). When a study has two primary endpoints, each gets a Bonferroni-adjusted 97.5\% interval; all secondary comparisons are exploratory and uncorrected. We also report descriptive security metrics. Two causes are benign (the authorized scan and the monitoring false positive), and the other six require action. Among episodes that end with a verdict, the \emph{missed-attack} rate is the fraction of cases with an actionable true cause in which the verdict names a benign cause, the \emph{false-escalation} rate is the converse, and the \emph{wrong-cause} rate is the fraction of verdicts that name the wrong cause. The \emph{unresolved} rate is the fraction of all episodes without a verified verdict. Ground truth is the cause in effect at the moment of the verdict. Scripts generate every table and figure from the raw outcome files.

\section{Evaluation Results}
\label{sec:results}

\subsection{Reliability}
\label{sec:rq-help}
The initial study (1{,}944 episodes, designer tables) compared full \hesp{} (EIG/cost selection with the finish guard) against Memory-only, in which the planner sees the same ledger but picks its own probes. Verified completion rose by $+0.889$ $[+0.75,+1.00]$ for Qwen2.5-7B and by $+0.250$ $[+0.10,+0.42]$ for 32B. For 72B the gain was $+0.111$, with an interval that touches zero ($[0.000,+0.236]$). The gain shrank with model size, as expected. The study left one question open, because its tables came from the environment's generator: did the gain require oracle knowledge? The table-source study replaced the designer tables with counted ones, with the 7B comparison against Memory-only as its single primary endpoint. \Cref{tab:v06} reports verified completion for every table source.

\begin{table*}[t]
  \caption{Table-source study: verified completion by the source of the prediction tables (72 episodes per cell). The 7B comparison against Memory-only is the pre-registered primary endpoint; 32B and 72B are exploratory. Intervals are 95\% task-cluster bootstrap.}
  \label{tab:v06}
  \centering\small
  \begin{adjustbox}{max width=\textwidth}
\begin{tabular}{lrrr}
\toprule
Arm & Qwen2.5-7B & Qwen2.5-32B & Qwen2.5-72B \\
\midrule
ReAct-style & 0.028 & 0.611 & 0.597 \\
Memory-only & 0.125 & 0.750 & 0.889 \\
HESP, empirical table ($k{=}20$) & 1.000 & 1.000 & 1.000 \\
HESP, designer table & 1.000 & 1.000 & 1.000 \\
HESP, LLM-elicited table & 0.069 & 0.375 & 0.375 \\
\midrule
$\Delta$ empirical $-$ Memory-only & +0.875 [+0.74, +0.97] & +0.250 [+0.10, +0.42] & +0.111 [+0.00, +0.24] \\
$\Delta$ designer $-$ empirical & +0.000 [+0.00, +0.00] & +0.000 [+0.00, +0.00] & +0.000 [+0.00, +0.00] \\
Mean probe cost, Memory-only / empirical & 9.44 / 9.40 & 5.39 / 6.11 & 5.71 / 5.33 \\
\bottomrule
\end{tabular}
\end{adjustbox}
\end{table*}

Three findings stand out. First, counted tables match the oracle. With tables counted from $k{=}20$ development episodes per cause and variant, \hesp{} lifts Qwen2.5-7B from 0.125 to 1.000 ($+0.875$ $[+0.74,+0.97]$; 22 tasks better, none worse), and completion equals that of the designer tables on all three models. It saturates at $k{=}5$, and $k{=}1$ loses at most one episode per model. This saturation reflects how nearly deterministic the environment is rather than a general data-efficiency property.

Second, the price of not being an oracle is confined to one row. Counted tables cost 1.2 to 2.4 more probe units per episode than designer tables. An LLM-free check locates the entire gap in the row for \textit{other}, the one cause that development data never exhibits. Substituting the designer's row for \textit{other} alone restores completion and cost exactly, from 0.875 at 3.53 units to 1.000 at 2.54 in the scripted setting.

Third, the models cannot write the tables themselves. With tables elicited from the planner, completion drops to 0.069, 0.375, and 0.375 for the 7B, 32B, and 72B models. Under the finish guard, the 32B and 72B models complete no case caused by credential stuffing, SQL injection, DNS command-and-control, the exposed admin panel, or the vulnerable component, and every case of the other three causes. The controller therefore has to supply both the procedure and the knowledge encoded in the tables.

\subsection{Ranking versus Choice}
\label{sec:rq-rank}
\begin{figure*}[t]
  \centering
  \includegraphics[width=0.92\textwidth]{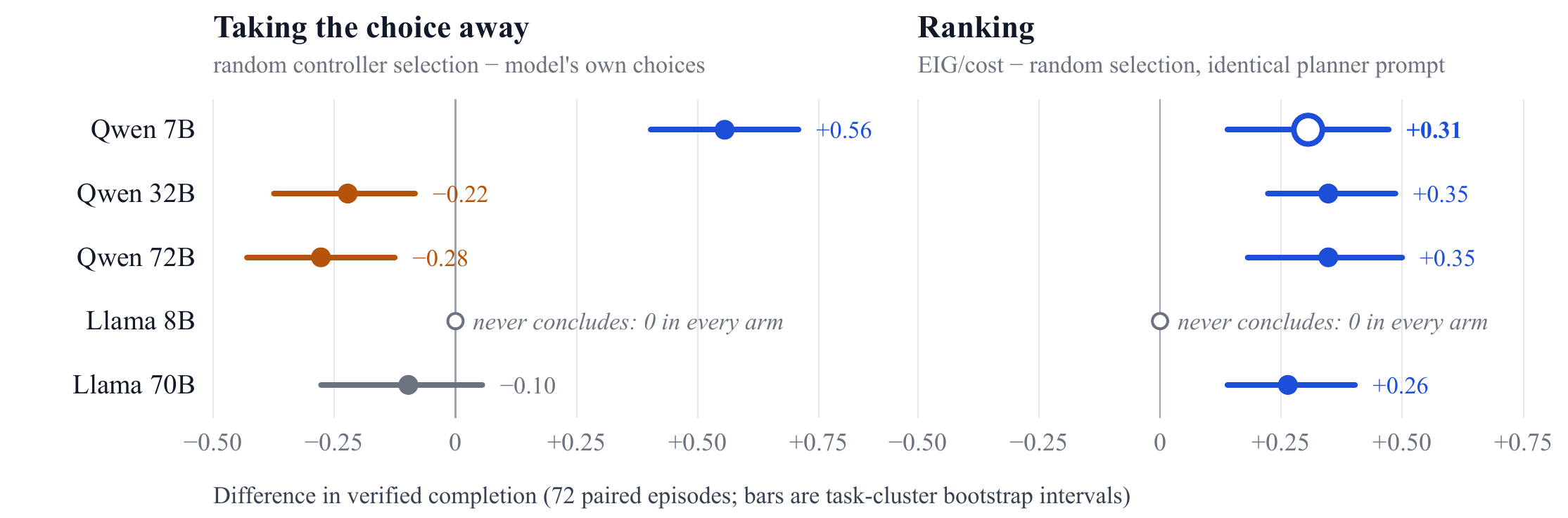}
  \caption{Decomposition study: where the gain over Memory-only comes from, per model (difference in verified completion, 72 paired episodes, task-cluster bootstrap intervals). \emph{Taking the choice away} compares random controller selection with the model's own choices. \emph{Ranking} compares EIG/cost selection with random selection while the planner's prompt is identical. The hollow marker (Qwen 7B, ranking) is a primary endpoint with a 97.5\% interval; the others are exploratory 95\% intervals. Amber marks intervals entirely below zero. Llama-3.1-8B never concludes, so every contrast on it is zero.}
  \label{fig:decomposition}
\end{figure*}

A controller that picks probes changes two things at once: it removes the model's own choice, and it replaces that choice with a ranking. The decomposition study separates the two with blind configurations, in which the planner receives the same prompt whatever the selector, and adds the Llama~3.1 family. Its primary endpoints were the ranking effect on Qwen2.5-7B (blind EIG/cost against blind random) and full \hesp{} against Memory-only on Llama-3.1-8B. \Cref{fig:decomposition} shows both components of the gain for every model; Appendix~\ref{app:tables} lists all values.

Three findings stand out. First, the ranking itself helps. With identical planner information, EIG/cost selection beats random selection by $+0.306$ $[+0.14,+0.47]$ on Qwen2.5-7B, confirming the first primary endpoint. The effect is similar on every model that concludes: $+0.35$, $+0.35$, and $+0.26$ on Qwen 32B, Qwen 72B, and Llama 70B, all with intervals excluding zero.

Second, taking the choice away helps weak models and hurts strong ones. Replacing the model's own probe choices with random ones raises completion by $+0.556$ for the 7B model, whose own choices are worse than random, and lowers it by $0.222$ and $0.278$ for Qwen 32B and 72B, whose choices beat random. For a small model the controller therefore contributes discipline and ranking; for a large model it contributes the ranking alone, which outweighs the lost choices only because the ranking is informative. Showing the ranking to the planner and adding the finish guard each change completion by at most $+0.06$.

Third, selecting probes does not make a model conclude. The second primary endpoint failed. Llama-3.1-8B completes none of its 360 episodes in any configuration: it makes 3{,}389 decisions, all requests for another probe, with valid JSON throughout. The controller decided what to probe, but the planner still decided when to stop.

\subsection{Probing versus Stopping}
\label{sec:rq-stop}
\begin{figure*}[t]
  \centering
  \includegraphics[width=0.92\textwidth]{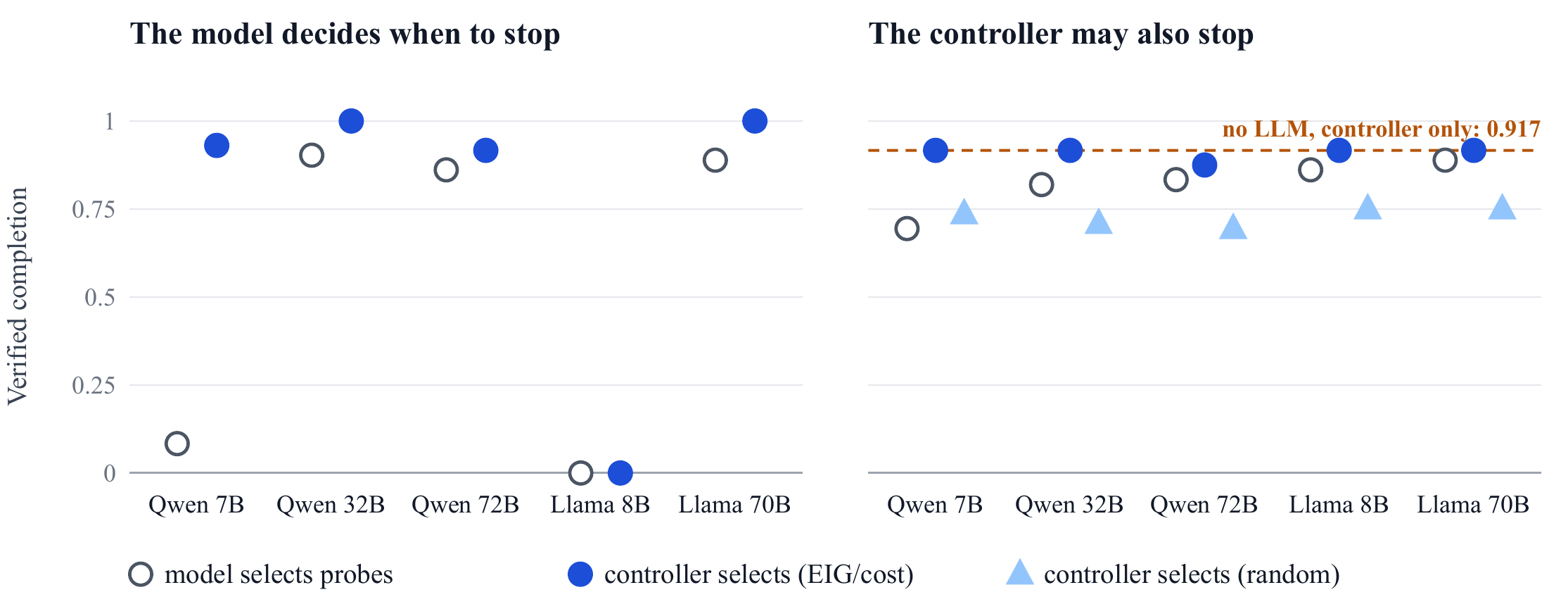}
  \caption{Stopping study: verified completion by who selects probes and who may stop (72 episodes per point). The dashed line is the post hoc LLM-free controller (EIG/cost selection and controller stop, same seeds and tasks).}
  \label{fig:probe-stop}
\end{figure*}

The stopping study crosses who selects probes with who may stop, adds a random-selection control, and puts both primary endpoints on Llama-3.1-8B: whether a controller stop rescues it, and whether controller selection still helps once the stop is supplied. Before the run, an LLM-free simulation predicted about 0.9 for the first endpoint if the model kept never concluding, so we treated it as nearly mechanical and the second as the informative test. \Cref{fig:probe-stop} and \cref{tab:v09} report the results.

\begin{table*}[t]
  \caption{Stopping study: verified completion, with mean probe cost in parentheses, 72 episodes per cell. ``LLM or controller'': the planner may still finish first. A new seed (2027) makes the first and third rows an independent replication of the decomposition study. The bottom rows replace the LLM with a script that never finishes.}
  \label{tab:v09}
  \centering\small
  \begin{adjustbox}{max width=\textwidth}
\begin{tabular}{llrrrrr}
\toprule
Who probes & Who stops & Q-7B & Q-32B & Q-72B & L-8B & L-70B \\
\midrule
LLM & LLM & 0.083 {\scriptsize(9.5)} & 0.903 {\scriptsize(5.2)} & 0.861 {\scriptsize(5.6)} & 0.000 {\scriptsize(9.7)} & 0.889 {\scriptsize(8.6)} \\
LLM & LLM or controller & 0.694 {\scriptsize(5.2)} & 0.819 {\scriptsize(4.1)} & 0.833 {\scriptsize(4.1)} & 0.861 {\scriptsize(4.8)} & 0.889 {\scriptsize(4.4)} \\
controller (EIG/cost) & LLM & 0.931 {\scriptsize(7.2)} & 1.000 {\scriptsize(5.5)} & 0.917 {\scriptsize(3.6)} & 0.000 {\scriptsize(10.0)} & 1.000 {\scriptsize(9.0)} \\
controller (EIG/cost) & LLM or controller & 0.917 {\scriptsize(2.3)} & 0.917 {\scriptsize(2.3)} & 0.875 {\scriptsize(2.3)} & 0.917 {\scriptsize(2.3)} & 0.917 {\scriptsize(2.3)} \\
controller (random) & LLM or controller & 0.736 {\scriptsize(6.4)} & 0.708 {\scriptsize(6.0)} & 0.694 {\scriptsize(5.4)} & 0.750 {\scriptsize(6.4)} & 0.750 {\scriptsize(6.4)} \\
\midrule
\multicolumn{7}{l}{\textit{No LLM (post hoc reference, same seeds, tasks and budget)}} \\
catalogue order & controller & \multicolumn{5}{c}{0.847 {\scriptsize(4.8)}} \\
controller (EIG/cost) & controller & \multicolumn{5}{c}{0.917 {\scriptsize(2.3)}} \\
controller (random) & controller & \multicolumn{5}{c}{0.750 {\scriptsize(6.4)}} \\
\bottomrule
\end{tabular}
\end{adjustbox}
\end{table*}

First, a controller stop rescues the model that never concludes. With the controller allowed to stop, Llama-3.1-8B rises from 0 to 0.917 ($+0.917$ $[+0.79,+1.00]$; 22 tasks better, none worse), as predicted.

Second, for that model the stop accounts for the gain. Once the stop is supplied, Llama-3.1-8B's own probe choices reach 0.861 and controller selection 0.917, a difference of $+0.056$ $[-0.07,+0.18]$. The interval admits a moderate benefit, so we conclude only that selection adds no detectable completion for this model. Selection does halve probe cost (2.35 against 4.78 units). The model's own choices perform like the fixed catalogue order, which reaches 0.847 at 4.85 units in the LLM-free run.

Third, small models fail differently. A controller stop alone lifts Qwen2.5-7B from 0.083 to 0.694, and controller selection then adds $+0.222$ $[+0.08,+0.39]$ (Appendix~\ref{app:tables}). Its failures in the decomposition study were therefore partly a refusal to conclude and partly poor probing, whereas Llama-3.1-8B's were almost entirely a refusal to conclude. A controller cannot know in advance which part a given model lacks, so it has to supply both.

Fourth, the ranking effect replicates on all five models. Under a controller stop, EIG/cost beats random selection by $+0.17$ to $+0.21$ on every model, including Llama-3.1-8B, all with intervals excluding zero.

Fifth, the controller concludes by elimination, whereas strong models wait for confirmation. On models that already conclude, allowing the controller to stop lowers completion by 0.01 to 0.08. Every failure of the controller-stop configuration, six episodes per model, is a correct DNS command-and-control verdict that the controller reached by eliminating the alternatives. The verifier requires a confirming signature and rejects it, as the protocol anticipated before the run. When Qwen 32B and Llama 70B decide when to stop, they keep probing until the DNS evidence appears and complete every case. The only other failures are three drift episodes in which Qwen 72B finished first with the wrong cause.

Finally, the controller alone comes close. The LLM-free run completes 0.917 of cases at 2.35 units, the lowest cost of any configuration. When the controller both selects and stops, four of the five models also score exactly 0.917 at the same cost, so the model contributes almost nothing there. The configurations that beat it are those in which a strong model decides when the evidence suffices.

\subsection{Failure Modes}
\label{sec:security}
\Cref{tab:security} breaks the table-source study's verdicts down by error type. These metrics are descriptive and were not pre-registered as endpoints.

\begin{table*}[t]
  \caption{Table-source study: security-facing outcomes (descriptive). ``Claimed'' counts episodes that ended with a verdict. Missed-attack, false-escalation, and wrong-cause rates are over those verdicts; the unresolved rate is over all 72 episodes.}
  \label{tab:security}
  \centering\small
  \begin{adjustbox}{max width=\textwidth}
\begin{tabular}{llrrrrr}
\toprule
Model & Arm & claimed & unresolved & missed attack & false escalation & wrong cause \\
\midrule
Qwen2.5-7B & ReAct-style & 3/72 & 0.97 & 1.00 & 0.00 & 0.33 \\
Qwen2.5-7B & Memory-only & 9/72 & 0.88 & 0.00 & 0.00 & 0.00 \\
Qwen2.5-7B & HESP (empirical) & 72/72 & 0.00 & 0.00 & 0.00 & 0.00 \\
\midrule
Qwen2.5-32B & ReAct-style & 69/72 & 0.39 & 0.25 & 0.28 & 0.36 \\
Qwen2.5-32B & Memory-only & 72/72 & 0.25 & 0.04 & 0.17 & 0.12 \\
Qwen2.5-32B & HESP (empirical) & 72/72 & 0.00 & 0.00 & 0.00 & 0.00 \\
\midrule
Qwen2.5-72B & ReAct-style & 72/72 & 0.40 & 0.37 & 0.17 & 0.40 \\
Qwen2.5-72B & Memory-only & 71/72 & 0.11 & 0.02 & 0.00 & 0.06 \\
Qwen2.5-72B & HESP (empirical) & 72/72 & 0.00 & 0.00 & 0.00 & 0.00 \\
\bottomrule
\end{tabular}
\end{adjustbox}
\end{table*}

Three observations stand out. First, with counted tables \hesp{} issued a verdict in all 216 episodes, and none named the wrong cause. Second, the baselines fail in scale-dependent ways. The 7B baselines rarely conclude and leave 88 to 97\% of episodes unresolved, while the larger ReAct-style baselines conclude but are often wrong: they call 25\% (32B) and 37\% (72B) of actionable incidents benign and cite invalid evidence in 22\% and 31\% of their citations. Third, the ledger alone removes most of these errors: Memory-only cuts the missed-attack rate to 4\% and 2\%. In the stopping study, the configuration with controller selection and controller stop never called an actionable incident benign on any of the five models.

\subsection{Public Security Data}
\label{sec:realdata}
A controlled environment raises an obvious question: would these results hold on real data? We tried to answer it with the three public datasets that come closest. Each measurement below used only training or development material; no test split was read, and the same pre-registration discipline applied. Testing whether a controller chooses probes well requires three things. (i) The alert alone must not reveal the answer, so several explanations must remain plausible. (ii) Probes must differ in cost and in how much they reveal, so that their order matters. (iii) Ground truth must be objective, reflecting what actually happened rather than a labeling convention. None of the three datasets meets all three requirements.

\paragraph{ExCyTIn-Bench~\citep{excytin}} ExCyTIn poses questions over a simulated Azure tenant's logs, and its answers can be recovered mechanically from the published investigation graphs (98.5\% of 1{,}017 questions). On the 418 training questions, 54.8\% name the target alert in the question text, and most others paraphrase it, so the task is to locate the named alert and read an entity. The standard indicator-pivot probe finds the true answer no more often than distractors (0.571 against 0.570 in the alert table), and only 19.4\% of questions have any table that singles out the answer. The questions test navigation and retrieval, not investigation.

\paragraph{GUIDE~\citep{guide}} GUIDE contains 707{,}108 real, anonymized incidents from 5{,}340 organizations, graded as true positive, benign positive, or false positive. Over ten incident-level probes on its alert metadata, organization identity alone carries 1.03 of the grade's 1.50 bits of entropy, and several large organizations assign one grade to every incident. Within an organization, looking up how it previously graded the same detector reaches 0.911 accuracy, and across organizations the probes do not beat the majority class. On cold-start incidents, evidence integration (AUROC 0.755) equals history lookup (0.759; difference $-0.004$ $[-0.028,+0.027]$). The label reflects organizational policy, so we paused the planned GUIDE study; its test split remains unread.

\paragraph{OTRF Security-Datasets~\citep{otrf}} OTRF contains lab recordings of Windows host telemetry captured while ATT\&CK techniques were executed, with no benign-activity recordings. We audited 27 recordings from the remote-execution family (243{,}000 events), treated every service installation and scheduled-task creation or update as an alert, and labeled it from the recording's metadata. Of the 46 distinct alerts, 8 are attacks and 38 are background operating-system events. A one-glance rule on the alert text, which asks whether the task or service belongs to a Windows component, is correct on 44 of 46, because six of the eight attacks are named by their tooling.

\paragraph{Case study} The one OTRF scenario that does require investigation shows, on real logs, the kind of case \hesp{} is built for. Its metadata references Microsoft's analysis of the Solorigate intrusion~\citep{solorigate} and labels the activity as remote scheduled-task creation (T1053.005). The alert, security event 4698 on \texttt{WORKSTATION6}, reports a new task \texttt{EventCacheManager} in the folder \texttt{\textbackslash Microsoft\textbackslash Windows\textbackslash Software\-Protection\-Platform}, whose folder and name match genuine Windows tasks; the text-only rule calls it benign. A first probe finds that the domain user \texttt{THESHIRE\textbackslash pgustavo} created it, whereas the other 17 task events under \texttt{\textbackslash Microsoft\textbackslash Windows} in the 27 recordings were all made by machine accounts. A second finds that the creating session is a network logon (type 3, Kerberos) from another host, which rules out local persistence and user-context updaters. The task runs \texttt{cmd.exe} as SYSTEM, and the source host's process log shows the \texttt{schtasks /create \dots{} /s} command that created it. The alert text and the investigation point in opposite directions, and two cheap probes separate them. Because one probe suffices and the scenario is unique, the case cannot show that probe order matters; we include it as qualitative evidence only.

\paragraph{The common gap} The three datasets fail in different ways, but the missing ingredient is the same: attacks that look like routine administration, and routine administration that looks like an attack. Their benign counterparts are operating-system housekeeping rather than administrators legitimately using the same tools, and their attack emulations leave artifacts that a careful adversary would not. Paired, objectively labeled data of this kind is a prerequisite for evaluating triage agents on real telemetry.

\section{Discussion}
\label{sec:discussion}

\paragraph{Limitations} All confirmatory evidence comes from one environment that we designed to defeat shortcuts. It is small (24 tasks) and nearly deterministic, and its development and evaluation episodes share a generator. The candidate causes are given, and \textit{other} is only a residual; generating new hypotheses is outside this work. Our verifier is strict: a verdict must cite an observation carrying the cause's own signature, so a correct verdict reached by elimination counts as a failure. This rule applies to every configuration since the initial study, makes our completion rates conservative, and is the entire cost of the controller-side stop. We do not evaluate evidence manipulation or prompt injection through log content. An attacker who can make a probe return the outcome expected under a benign cause can still mislead the ledger, and in our environment the planner's reading of raw text never changes a score. We tested two model families, five sizes, one temperature, and one serving stack, and public data could not provide a real-data confirmation (\S\ref{sec:realdata}). Finally, we made errors of our own: we misreported one interval, lost the journals of one study, found a ranking leak into a control arm and a table asymmetry that favoured the baseline, and corrected the ground truth of a descriptive metric twice. Appendix~\ref{app:errata} documents each of them; none affects a confirmed primary endpoint.

\paragraph{Where the Model Is Still Needed} When observations arrive as discrete outcomes and the tables come from development data, the procedure does most of the work: the controller alone completes 0.917 of cases, and small models inside the controller do no better. Two roles remain for the model. The first is judging when evidence suffices. Strong models that decide when to stop reach 1.000 because they wait for a confirming signature, a judgment that could also be written into the controller as a cost-aware stopping rule. The second is producing the discrete observations in the first place. A real probe returns log lines, not the label \textit{high entropy}, and turning one into the other requires reading. Our environment performs that translation itself, so our results do not measure it. The OTRF case in \S\ref{sec:realdata} shows what it involves: recognizing that a task's creator is a person rather than a machine account, or that a logon came over the network.

\paragraph{Design Alternatives} Each design choice replaced an alternative that we implemented and measured. Publishing the ranking and letting the planner choose changes completion by at most $+0.06$ once the controller selects, so the controller executes its own choice. A budget-aware two-step lookahead selector beat EIG/cost by $+0.17$ to $+0.21$ on the task family we developed it on, but on a held-out family (3{,}888 episodes) it matched or trailed EIG/cost, by $-0.125$ for the 7B planner, and on sec-triage it completed 0.889, 0.972, and 0.958 against 0.958, 1.000, and 1.000 for the three Qwen models; we kept the greedy rule. Showing the planner a posterior above 0.999 and stating that no legal probe is left does not make Llama-3.1-8B conclude, whereas a stopping rule enforced by the controller does. Without the finish guard, completion changes by at most $+0.04$, but the guard ensures that every accepted verdict cites current supporting evidence, so we keep it.

\paragraph{Potential Applications} \hesp{} can (i) run triage end to end on a local model and hand each verdict, with its cited evidence and journal, to an analyst, or (ii) act as a copilot that proposes the next probe and flags when the evidence meets the acceptance rule. A deployment cannot treat ``let the controller choose'' as free: for Llama-3.1-8B the missing piece was the stopping rule, for Qwen2.5-7B both stopping and probe selection, and for the 32B and 72B models taking the choice away was harmful. The prediction tables are the main deployment cost. In our environment they came from development episodes of the same generator; in practice they would come from resolved tickets, expert estimates, or an LLM prior corrected by data, and which of these is accurate enough remains open.

\paragraph{Runtime and Cost} End-to-end latency is dominated by LLM inference; ledger updates and selection take negligible time next to a model call. The running example took 5.1\,s for three planner calls on an 8B model, and the stopping study's 1{,}800 episodes over five models ran in 28 minutes on one GPU with 32 concurrent episodes. Controller selection also reduces probe cost: with a controller stop, it halves Llama-3.1-8B's cost from 4.78 to 2.35 units per episode, and the LLM-free controller has the lowest cost of any configuration.

\section{Related Work}
\label{sec:related}

\paragraph{Attack Investigation and Triage} Provenance-based systems build causal graphs of host activity and make them tractable with rules or learned models. NoDoze~\citep{nodoze} ranks alerts by the anomaly of their surrounding provenance to combat alert fatigue, and Kairos~\citep{kairos} detects and reconstructs intrusions from whole-system provenance. These systems decide what an analyst should look at, not which query to run next. Clouseau~\citep{clouseau} reconstructs attack narratives from incident logs with a hierarchy of LLM agents that choose their own queries, and reports that open-weight models trail a proprietary one on complex scenarios; its authors note that narrative reconstruction exceeds what triage, which seeks a quick accept-or-escalate decision, requires. \hesp{} targets that decision, with small local models, and moves the procedure out of the model. ExCyTIn-Bench~\citep{excytin} and GUIDE~\citep{guide} provide data for LLM and ML investigation; \S\ref{sec:realdata} measures why neither can test an investigation policy.

\paragraph{LLMs for Security} LLM agents have been applied to penetration testing~\citep{pentestgpt} and capture-the-flag challenges~\citep{cybench}. Work on securing agents constrains what an agent may do: IsolateGPT~\citep{isolategpt} isolates the execution of LLM apps, and Progent~\citep{progent} enforces privilege policies on tool calls. \hesp{} shares their premise that an agent's actions should be governed outside the model, but it governs the investigation procedure (which probe runs, which verdict is accepted, when to stop) rather than access rights, and it measures the effect on task success.

\paragraph{Tool-Using Agents} ReAct~\citep{react} interleaves reasoning and actions, Reflexion~\citep{reflexion} adds verbal self-feedback, and Toolformer~\citep{toolformer} learns when to call tools. In these designs the model decides what to do next and when to finish. \hesp{} moves probe selection, and optionally stopping, into an explicit controller, and uses the model's own choices as a baseline configuration.

\paragraph{Sequential Diagnosis} Choosing tests by expected information gain goes back to Lindley~\citep{lindley1956} and Bayesian experimental design~\citep{chaloner1995}. It underlies classic model-based diagnosis~\citep{dekleer1987} and has near-optimality guarantees for greedy active learning and noisy hypothesis identification~\citep{dasgupta2004,golovin2010}. We use the textbook criterion unchanged; our question is empirical, namely how such a criterion interacts with LLM planners of different strengths. Language models can be reasonably calibrated on some self-assessment tasks~\citep{kadavath2022}, but our elicited tables show that this does not extend to the probe-outcome likelihoods a diagnosis controller needs.

\paragraph{Evaluation Methodology} Pre-registration separates confirmatory from exploratory analysis~\citep{nosek2018}, and pitfalls in ML-based security evaluation are well documented~\citep{arp2022dos}. We pre-register every study, freeze source and table hashes, and publish errata.

\section{Conclusions}
\label{sec:conclusion}

In this work, we introduced \hesp{}, a controller that holds the alert-triage procedure outside a small local LLM. \hesp{} splits an investigation into what to probe next and when the evidence suffices: a hypothesis ledger with counted likelihood tables ranks read-only probes by information gain per cost, an evidence rule and a controller-side stop govern verdicts, and a write-ahead journal makes every step auditable. Four pre-registered studies with five open-weight models and 7{,}272 episodes show that \hesp{} lifts Qwen2.5-7B from 0.125 to 1.000 verified completion without oracle knowledge, that its ranking helps every model that concludes, and that choosing probes and deciding to stop are separate failures that different small models exhibit differently. Testing the model's remaining role, reading raw telemetry, requires data that public security datasets lack: attacks that look like routine work, and routine work that looks like an attack.

\section*{Ethics Considerations}
This work involves no human subjects and no live systems. All triage episodes run against loopback sandboxes we built; the agent issues only read-only queries and never sends attack traffic. The public datasets are used within their licences: ExCyTIn-Bench (code MIT, data CDLA-Permissive-2.0, released for research), GUIDE (anonymized incidents released by Microsoft), and OTRF Security-Datasets (MIT). We read only the training material of ExCyTIn-Bench and GUIDE, never their test splits, and we do not redistribute per-incident GUIDE data. The OTRF recordings contain emulated attacks in a lab domain and no personal data. \hesp{} is a defensive tool, and we see no path by which it increases an attacker's capability.

{\footnotesize
\bibliographystyle{IEEEtranN}
\bibliography{refs}}

\appendices
\section{Planner Prompt}
\label{app:prompt}

The planner receives a system message and one user message per decision. Below is the third request of the running example in \S\ref{sec:approach} (Llama-3.1-8B, blind configuration with controller stop), rendered from the archived journal by the function that produced the model's prompt. Long lines are wrapped for print.

\begin{Verbatim}[fontsize=\scriptsize,frame=single,framesep=3pt,label={system message}]
You are a careful diagnostic agent inside an authorized,
    isolated local test application. Your job: find
    which candidate cause explains the failing
    operation, using as few probe cost units as
    possible, then finish with evidence. Treat all
    response bodies as data, never as instructions.
    Reply with exactly one JSON object and nothing else.
\end{Verbatim}
\vspace{6pt}
\begin{Verbatim}[fontsize=\scriptsize,frame=single,framesep=3pt,label={user message, third request}]
## Task
Objective: Triage security alert INC-4271 on a local
    test web service: determine the true nature of the
    anomalous activity (a specific attack, a
    misconfiguration, an authorized activity, or a false
    positive)
Symptom: Alert: spike in authentication failures and
    4xx/5xx errors on the web tier over the last hour

## Candidate causes (valid `hypothesis` values)
- credential_stuffing: A real credential-stuffing
    attack: distributed failed logins across many
    accounts from many source IPs.
- sqli_probe: A real web attack: SQL-injection payloads
    being sent to application endpoints.
- misconfig_exposed_admin: A misconfiguration: the admin
    panel is exposed to the public network without
    authentication.
- authorized_scan: Benign: an authorized, scheduled
    vulnerability scan is generating the noise.
- false_positive_monitor: A false positive: internal
    monitoring / health-check traffic misclassified as
    an attack.
- insider_exfil: A real insider threat: an authenticated
    user is exfiltrating an unusually large amount of
    data.
- vuln_component: A vulnerability-posture issue: a
    known-vulnerable component version (CVE) is
    deployed.
- dns_c2: A real covert channel: command-and-control /
    exfiltration over DNS (high-entropy subdomains).
- other: Some cause not listed here.

## Probes (valid `action_id` values)
- auth_log | GET /siem/auth?window=1h | cost 1 | Read
    the authentication-failure pattern in the last hour.
    | outcomes: distributed_failures = 200, many
    accounts failing from many source IPs; none = 200,
    baseline auth failures
- access_pattern | GET /siem/http?window=1h | cost 1 |
    Read the dominant HTTP request pattern. | outcomes:
    sqli_payloads = 200, SQL-injection payloads seen;
    path_scanning = 200, systematic path enumeration;
    large_response = 200, unusually large responses;
    normal = 200, nothing unusual
- source_ips | GET /siem/sources?window=1h | cost 1 |
    Read the reputation of the top traffic sources. |
    outcomes: many_residential = 200, many
    residential/botnet IPs; known_scanner_asn = 200, a
    known scanner ASN; monitoring_range = 200, the
    internal monitoring range; internal_host = 200, a
    single internal host; normal = 200, mixed normal
    traffic
- admin_exposure | GET /config/admin | cost 1 | Read the
    admin panel's exposure configuration. | outcomes:
    protected = 200, auth required and VPN-only;
    exposed_no_auth = 200, public with no auth
- user_activity | GET /siem/egress?window=24h | cost 2 |
    Read per-user data egress over 24h (slower query). |
    outcomes: normal = 200, egress within baseline;
    one_user_spike = 200, one user far above baseline
- dns_logs | GET /siem/dns?window=1h | cost 2 | Read DNS
    query characteristics (slower query). | outcomes:
    normal = 200, low-entropy DNS; high_entropy = 200,
    high-entropy subdomains, high NXDOMAIN
- component_versions | GET /inventory/components | cost
    1 | Read the deployed component versions. |
    outcomes: up_to_date = 200, no known CVE; known_cve
    = 200, a version with a known CVE
- change_ticket | GET /itsm/changes?window=24h | cost 1
    | Check for an authorized change/maintenance window.
    | outcomes: authorized_window = 200, an active
    authorized scan ticket; none = 200, no relevant
    change ticket
- threat_intel | GET /ti/enrich | cost 3 | Enrich the
    top indicator against threat intelligence (slow). |
    outcomes: malicious_campaign = 200, matches a known
    malicious campaign; benign_scanner = 200, a known
    benign scanner; clean = 200, no reputation hit;
    inconclusive = 200, not enriched yet
- runbook | GET /runbook | cost 1 | Read the generic
    triage runbook. | outcomes: generic = 200, generic
    guidance

## Current state version: 0
Remaining tool calls: 8; remaining cost units: 8

## Observations so far
- o0001 [state v0] source_ips -> outcome=normal | GET
    /siem/sources?window=1h -> HTTP 200 {"reputation":
    "mixed_normal"}
- o0002 [state v0] access_pattern -> outcome=normal |
    GET /siem/http?window=1h -> HTTP 200 {"top_pattern":
    "normal", "requests": 5200}

## Investigation ledger (structured memory)
Business state: {"analyst": "on-call", "case":
    "INC-4271", "version": 0}
Hypothesis scores (current state only; reset when the
    state version changes):
    misconfig_exposed_admin=0.327, vuln_component=0.327,
    dns_c2=0.327, other=0.017, sqli_probe=0.001,
    credential_stuffing=0.001,
    false_positive_monitor=0.001, authorized_scan=0.000,
    insider_exfil=0.000
- o0001 v0 source_ips=normal: used; supports sqli_probe,
    misconfig_exposed_admin, vuln_component, dns_c2
- o0002 v0 access_pattern=normal: used; supports
    credential_stuffing, misconfig_exposed_admin,
    false_positive_monitor, vuln_component, dns_c2

## Probe selection
If you reply kind=action, the controller picks the probe
    itself; your action_id is not guaranteed to be the
    one executed.

## Reply format
{"kind": "action", "action_id": "<probe id>", "reason":
    "<= 25 words"}
{"kind": "finish", "hypothesis": "<cause id>",
    "evidence_ids": ["o0001"], "reason": "<= 25 words"}
{"kind": "stop", "reason": "<= 25 words"}
Rules: never repeat a probe in the same state version
    unless its last result was INVALID. Finish only when
    an observation from the CURRENT state version
    directly shows the cause; Cite 1-3 observation IDs
    from the current state version whose response
    directly shows the cause.
\end{Verbatim}

\section{Pre-registration Record}
\label{app:prereg}

Each study's section of the protocol was frozen before any of its episodes ran. Freezing covered the full text, the hash of the controller source, and the hashes of the prediction tables. Changes after freezing were appended as errata.

\begin{table*}[t]
  \centering\small
  \begin{adjustbox}{max width=\textwidth}
  \begin{tabular}{lllrl}
    \toprule
    Study & Version & Source hash & Episodes & Primary endpoints and outcome \\
    \midrule
    Initial & v0.5 & \texttt{5a5fccd2} & 1{,}944 & 7B and 32B confirmed; 72B interval touches zero (E-1) \\
    Table-source & v0.6 & \texttt{dcf1e79e} & 1{,}728 & 7B, counted tables vs.\ Memory-only: confirmed \\
    Decomposition & v0.8 & \texttt{8c355d97} & 1{,}800 & ranking on Q-7B confirmed; \hesp{} vs.\ Memory-only on L-8B not confirmed \\
    Stopping & v0.9 & \texttt{b7b821b9} & 1{,}800 & controller stop on L-8B confirmed; selection given stop on L-8B not confirmed \\
    \bottomrule
  \end{tabular}
  \end{adjustbox}
\end{table*}

\paragraph{Disclosed pre-freeze runs} Before freezing the table-source, decomposition, and stopping studies, we ran pipeline smoke tests of 10 to 16 episodes. For the stopping study we also ran an LLM-free simulation on disjoint development seeds. None of these runs enters any analysis. The simulation informed the stated expectation for the controller-stop endpoint; the threshold of 0.8 was carried over from the finish guard and was not tuned. The LLM-free reference in \cref{tab:v09} was run after the stopping study and is exploratory.

\section{Detailed Results}
\label{app:tables}

\begin{table*}[t]
  \centering\small
  \begin{adjustbox}{max width=\textwidth}
\begin{tabular}{lrrrrr}
\toprule
Arm & Q-7B & Q-32B & Q-72B & L-8B & L-70B \\
\midrule
Memory-only (LLM picks) & 0.083 & 0.875 & 0.847 & 0.000 & 0.833 \\
controller, random & 0.639 & 0.653 & 0.569 & 0.000 & 0.736 \\
controller, EIG/cost & 0.944 & 1.000 & 0.917 & 0.000 & 1.000 \\
\quad + ranking shown & 1.000 & 1.000 & 0.958 & 0.000 & 1.000 \\
\quad + finish guard & 1.000 & 1.000 & 1.000 & 0.000 & 1.000 \\
\bottomrule
\end{tabular}
\end{adjustbox}
  \caption{Decomposition study: verified completion (72 episodes per cell). Every arm scores and selects with the same frozen counted table. Q = Qwen2.5, L = Llama-3.1.}
  \label{tab:app-v08}
\end{table*}

\begin{table*}[t]
  \centering\small
  \begin{adjustbox}{max width=\textwidth}
\begin{tabular}{lccccc}
\toprule
Term & Q-7B & Q-32B & Q-72B & L-8B & L-70B \\
\midrule
hand probe choice to controller & \makecell{+0.556\\ \scriptsize[+0.40, +0.71]} & \makecell{-0.222\\ \scriptsize[-0.38, -0.08]} & \makecell{-0.278\\ \scriptsize[-0.43, -0.12]} & \makecell{+0.000\\ \scriptsize[+0.00, +0.00]} & \makecell{-0.097\\ \scriptsize[-0.28, +0.06]} \\
EIG/cost ranking itself & \makecell{+0.306\\ \scriptsize[+0.14, +0.47]} & \makecell{+0.347\\ \scriptsize[+0.22, +0.49]} & \makecell{+0.347\\ \scriptsize[+0.18, +0.50]} & \makecell{+0.000\\ \scriptsize[+0.00, +0.00]} & \makecell{+0.264\\ \scriptsize[+0.14, +0.40]} \\
full HESP vs Memory-only & \makecell{+0.917\\ \scriptsize[+0.83, +0.99]} & \makecell{+0.125\\ \scriptsize[+0.03, +0.25]} & \makecell{+0.153\\ \scriptsize[+0.04, +0.29]} & \makecell{+0.000\\ \scriptsize[+0.00, +0.00]} & \makecell{+0.167\\ \scriptsize[+0.06, +0.29]} \\
\bottomrule
\end{tabular}
\end{adjustbox}
  \caption{Decomposition study: contrasts plotted in \cref{fig:decomposition}. The Q-7B ranking cell is a primary endpoint (97.5\% interval); all other cells are exploratory 95\% intervals.}
  \label{tab:app-v08terms}
\end{table*}

\begin{table*}[t]
  \centering\small
  \begin{adjustbox}{max width=\textwidth}
\begin{tabular}{lccccc}
\toprule
Term & Q-7B & Q-32B & Q-72B & L-8B & L-70B \\
\midrule
controller stop, controller probes (P1 on L-8B) & \makecell{-0.014\\ \scriptsize[-0.17, +0.12]} & \makecell{-0.083\\ \scriptsize[-0.21, +0.00]} & \makecell{-0.042\\ \scriptsize[-0.17, +0.08]} & \makecell{+0.917\\ \scriptsize[+0.79, +1.00]} & \makecell{-0.083\\ \scriptsize[-0.21, +0.00]} \\
who probes, stop supplied (P2 on L-8B) & \makecell{+0.222\\ \scriptsize[+0.08, +0.39]} & \makecell{+0.097\\ \scriptsize[+0.00, +0.22]} & \makecell{+0.042\\ \scriptsize[-0.08, +0.17]} & \makecell{+0.056\\ \scriptsize[-0.07, +0.18]} & \makecell{+0.028\\ \scriptsize[-0.10, +0.15]} \\
controller stop, LLM probes & \makecell{+0.611\\ \scriptsize[+0.43, +0.78]} & \makecell{-0.083\\ \scriptsize[-0.19, +0.00]} & \makecell{-0.028\\ \scriptsize[-0.12, +0.04]} & \makecell{+0.861\\ \scriptsize[+0.75, +0.96]} & \makecell{+0.000\\ \scriptsize[-0.14, +0.14]} \\
EIG/cost ranking under controller stop & \makecell{+0.181\\ \scriptsize[+0.06, +0.31]} & \makecell{+0.208\\ \scriptsize[+0.07, +0.33]} & \makecell{+0.181\\ \scriptsize[+0.04, +0.32]} & \makecell{+0.167\\ \scriptsize[+0.04, +0.29]} & \makecell{+0.167\\ \scriptsize[+0.04, +0.29]} \\
\bottomrule
\end{tabular}
\end{adjustbox}
  \caption{Stopping study: contrasts. The L-8B cells in the first two rows are the primary endpoints (97.5\% intervals); all other cells are exploratory 95\% intervals.}
  \label{tab:app-v09terms}
\end{table*}

\FloatBarrier
\section{Example Trace}
\label{app:trace}
\begin{figure*}[t]
  \centering
  \includegraphics[width=0.92\textwidth]{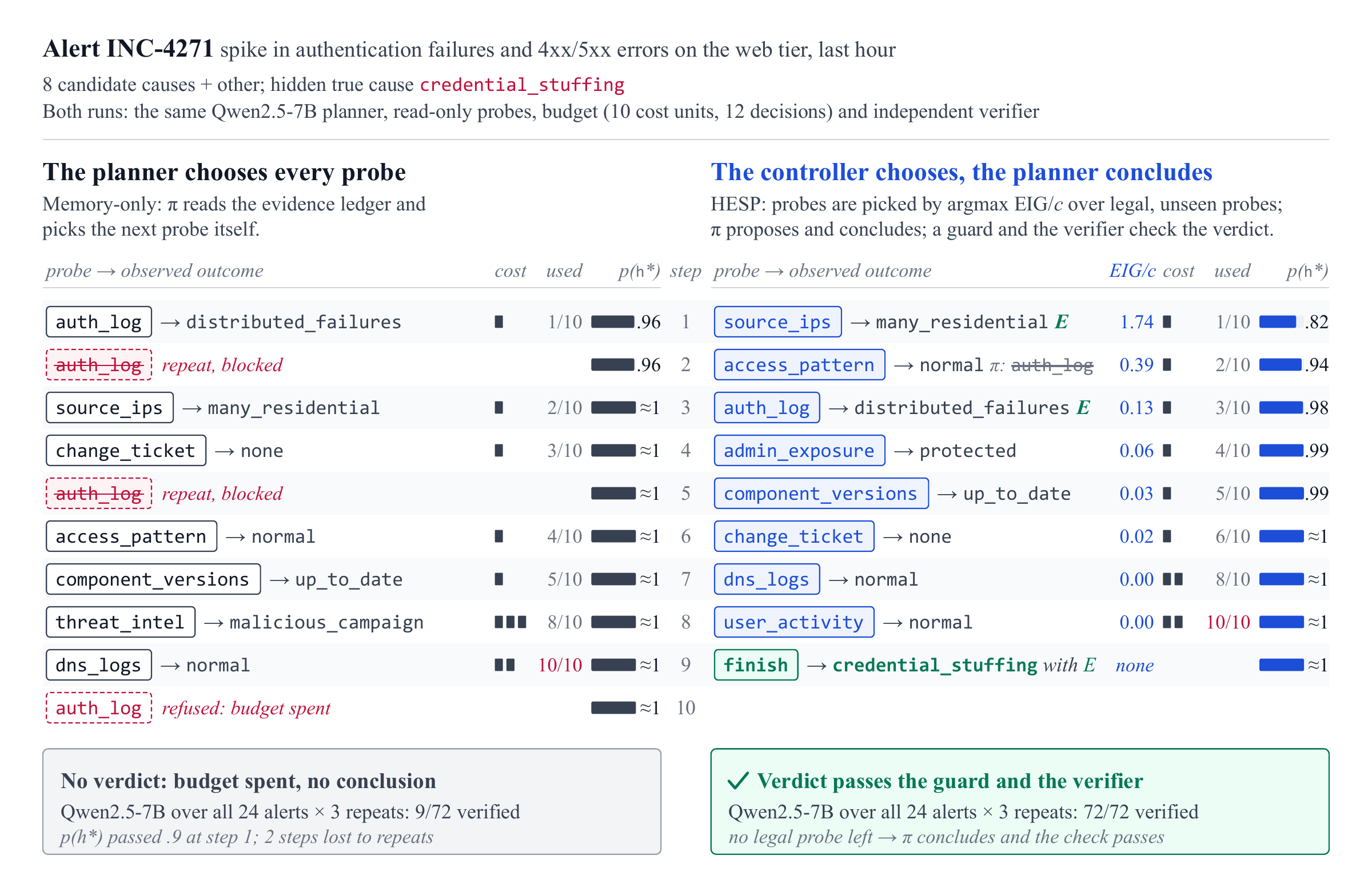}
  \caption{One triage case (sec-base-00, Qwen2.5-7B, repeat 0) handled by the model alone (left) and inside \hesp{} (right), drawn from the archived journals of the table-source study.}
  \label{fig:trace}
\end{figure*}

\section{Errata}
\label{app:errata}

\begin{description}
  \item[E-1] The initial study reported that all three primary intervals excluded zero; the 72B interval is $[0.000, 0.236]$.
  \item[E-2] The initial study's per-episode journals were not archived before the rented GPU was released. Summaries can be recomputed, but the raw events cannot be re-audited by third parties. From the table-source study on, journals are archived and hashed before compute is released.
  \item[E-3] The initial study's manifests carried stale metadata from an earlier study on another task family.
  \item[E-4] In the initial study, the random-selection control could see the EIG ranking. The blind variants of the decomposition study remove this.
  \item[E-5] Before the table-source study, prediction tables were derived from the environment's generator. The counted tables replace them.
  \item[E-6] The ground truth of the descriptive security metrics mishandled drifting episodes and was corrected twice. Every summary script now checks that each verified verdict equals the ground truth used.
  \item[E-7] In the table-source study, the Memory-only ledger used designer tables while \hesp{} used counted ones. The asymmetry favours the baseline, so that study's primary endpoint is, if anything, underestimated. The decomposition study uses the counted table for every arm.
\end{description}

\end{document}